\documentclass[12pt]{iopart}

\usepackage{graphicx}
\begin{document}

\title{A NLO analysis on fragility of
dihadron tomography in high energy $AA$ collisions }

\author{Hanzhong Zhang$^1$, J.~F. Owens$^2$, Enke Wang$^1$ and Xin-Nian Wang$^3$}

\address{$^1$Institute of Particle Physics, Huazhong Normal University,
         Wuhan 430079, China}
\address{$^2$Physics Department, Florida State University, Tallahassee,
          FL 32306-4350, USA}
\address{$^3$Nuclear Science Division, Lawrence Berkeley Laboratory,
         Berkeley, CA 94720, USA}
\ead{zhanghz@iopp.ccnu.edu.cn}

\begin{abstract}
The dihadron spectra in high energy $AA$ collisions are studied
within the NLO pQCD parton model with jet quenching taken into
account. The high $p_T$ dihadron spectra are found to be
contributed not only by jet pairs close and tangential to the
surface of the dense matter but also by punching-through jets
survived at the center while the single hadron high $p_T$ spectra
are only dominated by surface emission. Consequently, the
suppression factor of such high-$p_T$ hadron pairs is found to be
more sensitive to the initial gluon density than the single hadron
suppression factor.
\end{abstract}

\begin{figure}[tb]
\hspace{.2in}
\includegraphics[width=0.5\linewidth]{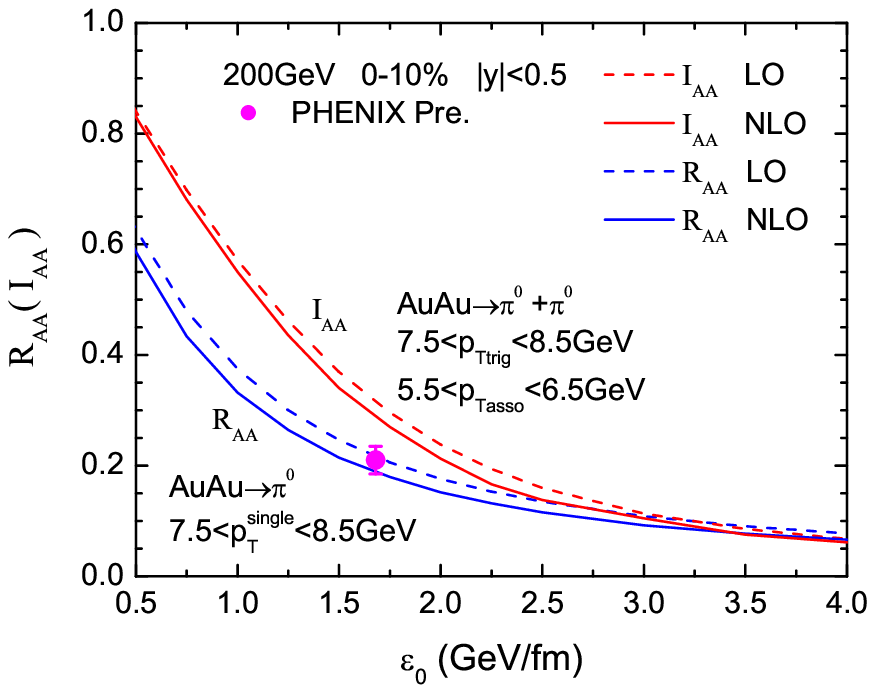}
\vskip -2.55in\hspace{3.3in}
\includegraphics[width=0.5\linewidth]{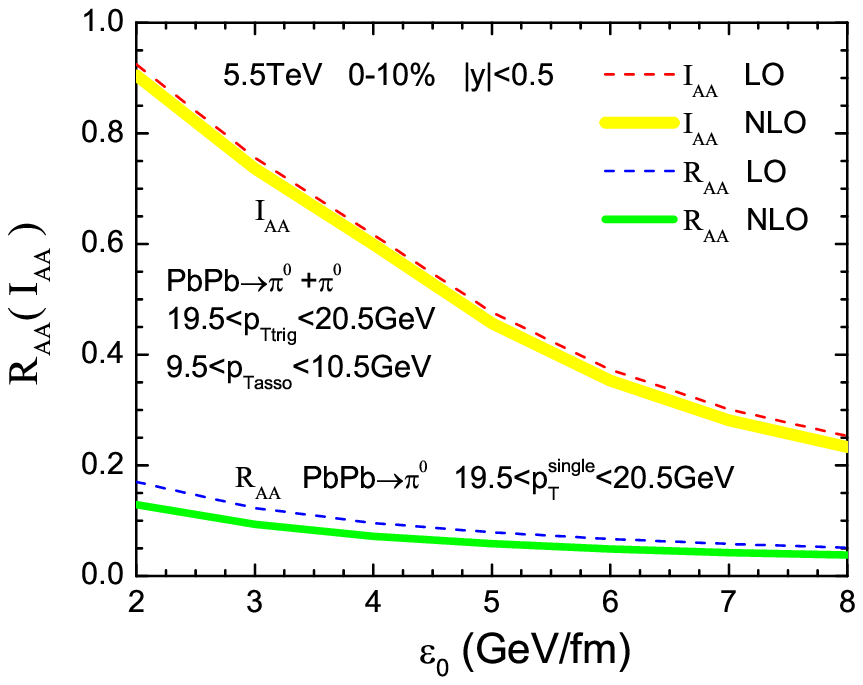}
\vskip -.15in \caption{\label{fig:RaaIaa} The suppression factors
for single ($R_{AA}$)
 and dihadron ($I_{AA}$) spectra as a function of the initial energy loss
parameter $\epsilon_0$. The data are from
Ref.\protect\cite{phenixnnaapt}.}
\end{figure}

One of the most exciting phenomena observed\cite{rhic0203} at the
Relativistic Heavy ion Collider (RHIC) is jet
quenching\cite{Wang94}---a hard probe of a strongly-interacting
quark gluon plasma in high energy heavy ion collisions. The
observed suppression of large $p_T$ hadron spectrum is caused by
the total parton energy loss which is related to the average gluon
density along the jet propagation path and the total propagation
length\cite{xn04}. Therefore, measurements of large $p_T$ hadron
suppression can be directly related to the averaged gluon density.
Here we will employ a NLO pQCD parton model\cite{owens02} to study
the suppression of both single and dihadron spectra due to jet
quenching. Different from the previous LO study\cite{xn04},
because the number ratio of gluon/quark jets is larger in NLO than
in LO calculation and the energy loss of a gluon jet is assumed to
be 9/4 larger than that of a quark jet, NLO contribution will
behave with stronger quenching effect than LO contribution (see
Fig.~\ref{fig:RaaIaa}, $R_{AA}^{NLO}<R_{AA}^{LO}$,
$I_{AA}^{NLO}<I_{AA}^{LO}$). In particular, we will check the
robustness of back-to-back dihadron spectra as a probe of the
initial gluon density when single hadron spectra supression become
fragile\cite{zoww}.

Within a NLO pQCD parton model \cite{owens02}, large $p_T$
particle production cross section in $N+N$ collisions can be
expressed as a convolution of NLO parton-parton scattering cross
sections, parton distributions inside the collided nucleons and
parton fragmentation functions (FF). In order to study large $p_T$
particle production in $A+A$ collisions, one can extrapolate $N+N$
cross section to $A+A$ collisions.  The effect of jet quenching in
$A+A$ collisions is incorporated via the modified jet
fragmentation functions due to radiative parton energy loss in
dense medium \cite{xn04,WW}. The modified jet fragmentation
functions are in turn characterized by the average radiative
parton energy loss which is proportional to the initial gluon
density. An energy loss parameter $\epsilon_0$ \cite{xn04,WW} is
introduced in following numerical calculations, which is
proportional to the initial gluon density $\rho_0$.

\begin{figure}[tb]
\hspace{.2in}
\includegraphics[width=0.5\linewidth]{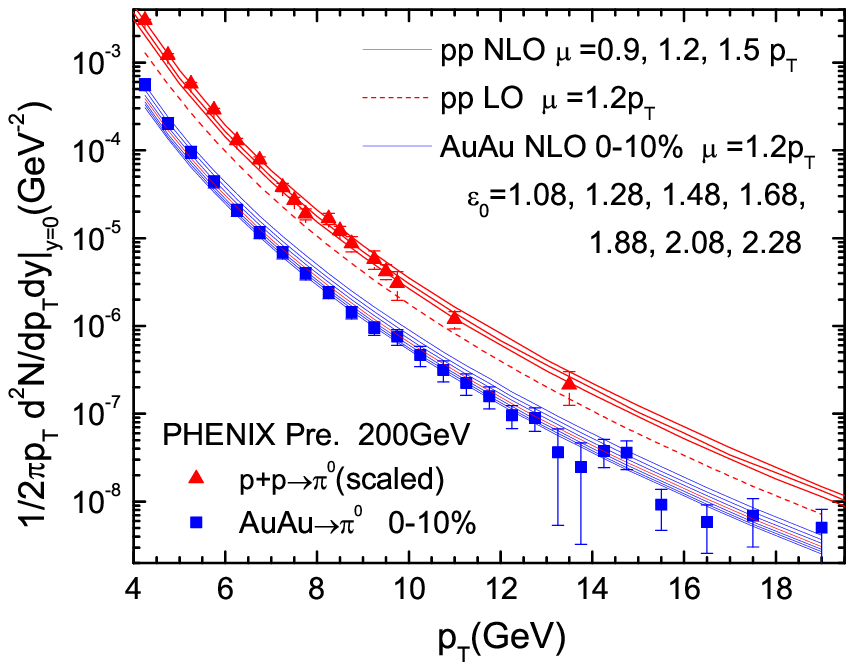}
\vskip -2.55in\hspace{3.3in}
\includegraphics[width=0.5\linewidth]{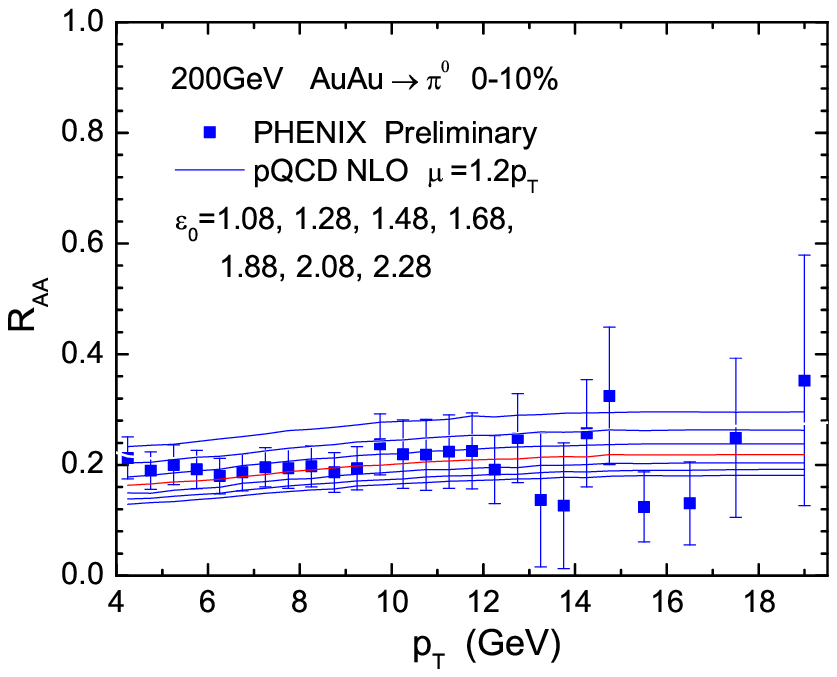}
\vskip -.15in \caption{\label{fig:ppRaa} The single $\pi^0$ spectra
 (left) in $p+p$ and in central $Au+Au$ collisions,
 and the nuclear modification factors (right). The data are from
Ref.\protect\cite{phenixnnaapt}.}
\end{figure}

\begin{figure}[tb]
\hspace{.2in}
\includegraphics[width=0.5\linewidth]{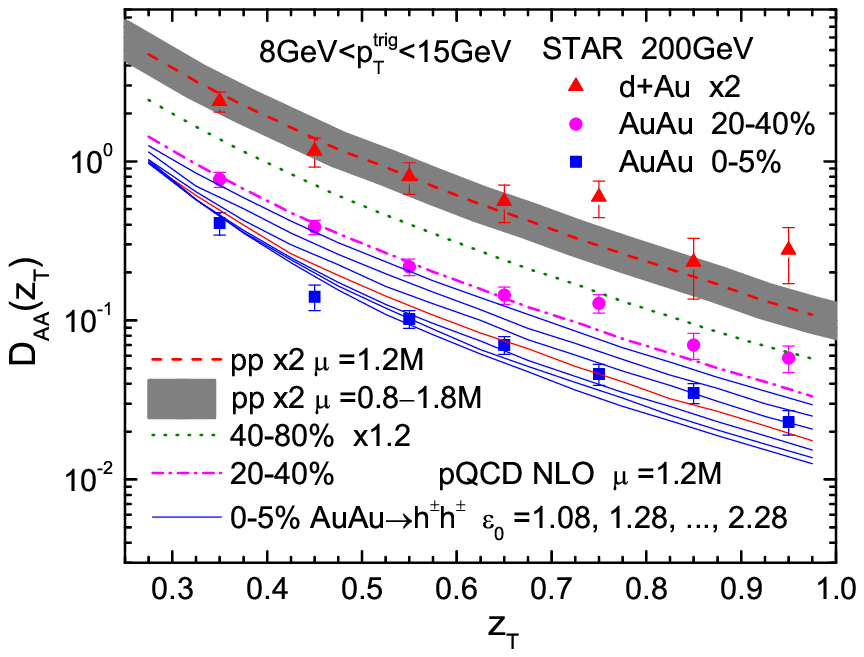}
\vskip -2.55in\hspace{3.3in}
\includegraphics[width=0.5\linewidth]{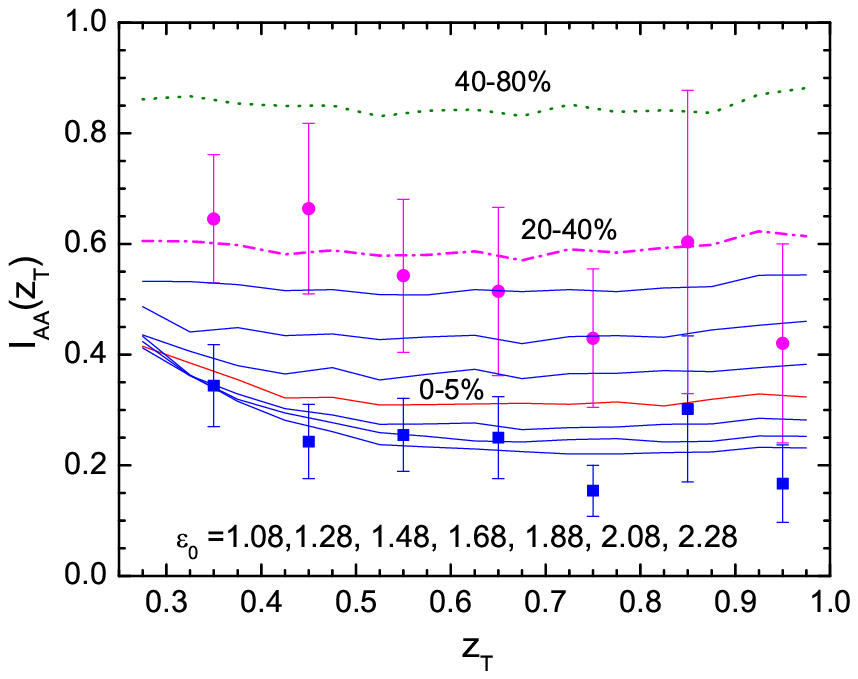}
\vskip -.15in \caption{\label{fig:DaaIaa} The associated hadron
spectra (left) of dihadron
 in $p+p$ and $Au+Au$ collisions,
 and the suppression factor (right) of dihadron. The data are from
Ref.\protect\cite{STAR-dijet06}.}
\end{figure}

\begin{figure}[tb]
\hspace{.2in}
\includegraphics[width=0.45\linewidth]{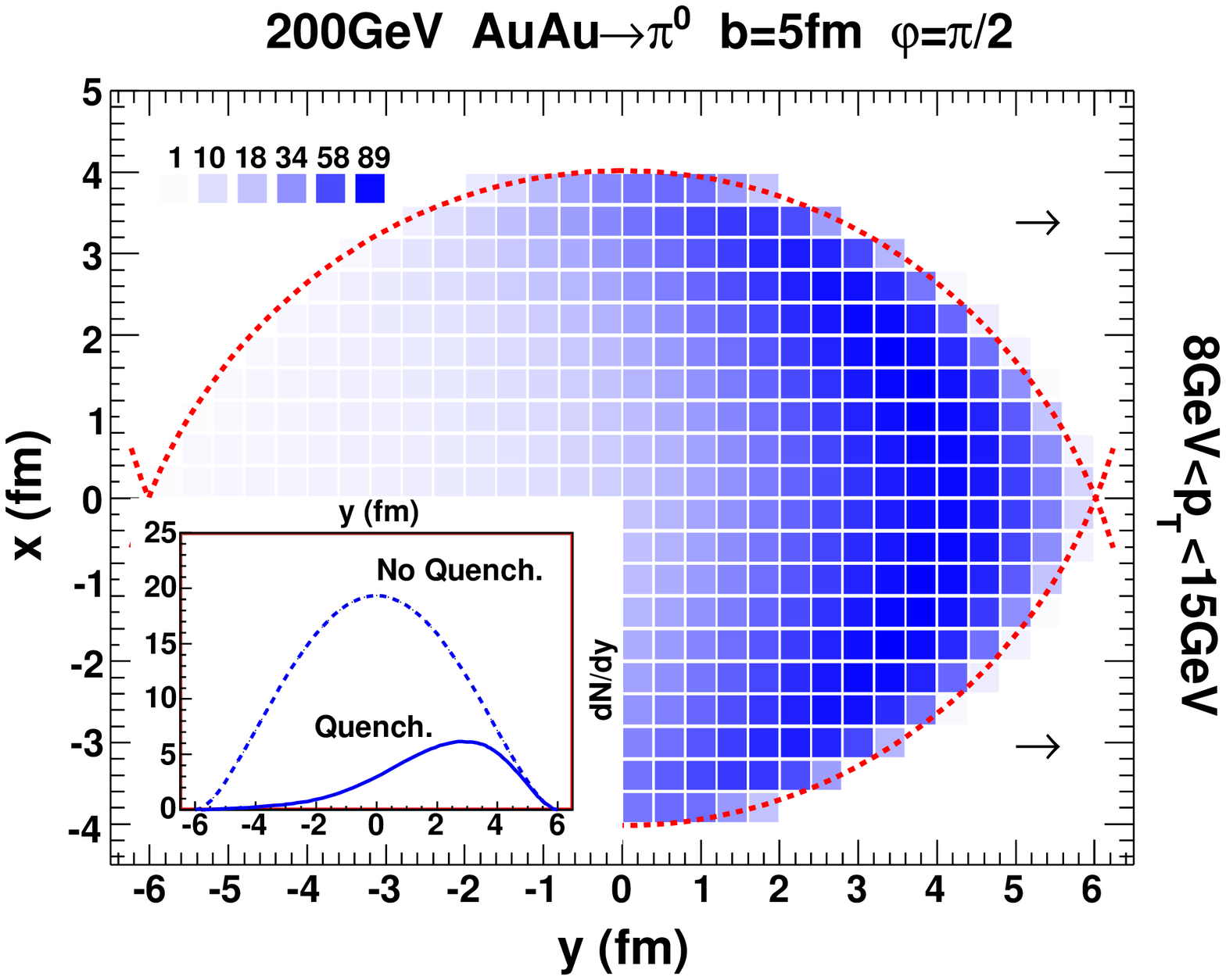}
\vskip -2.25in\hspace{3.3in}
\includegraphics[width=0.45\linewidth]{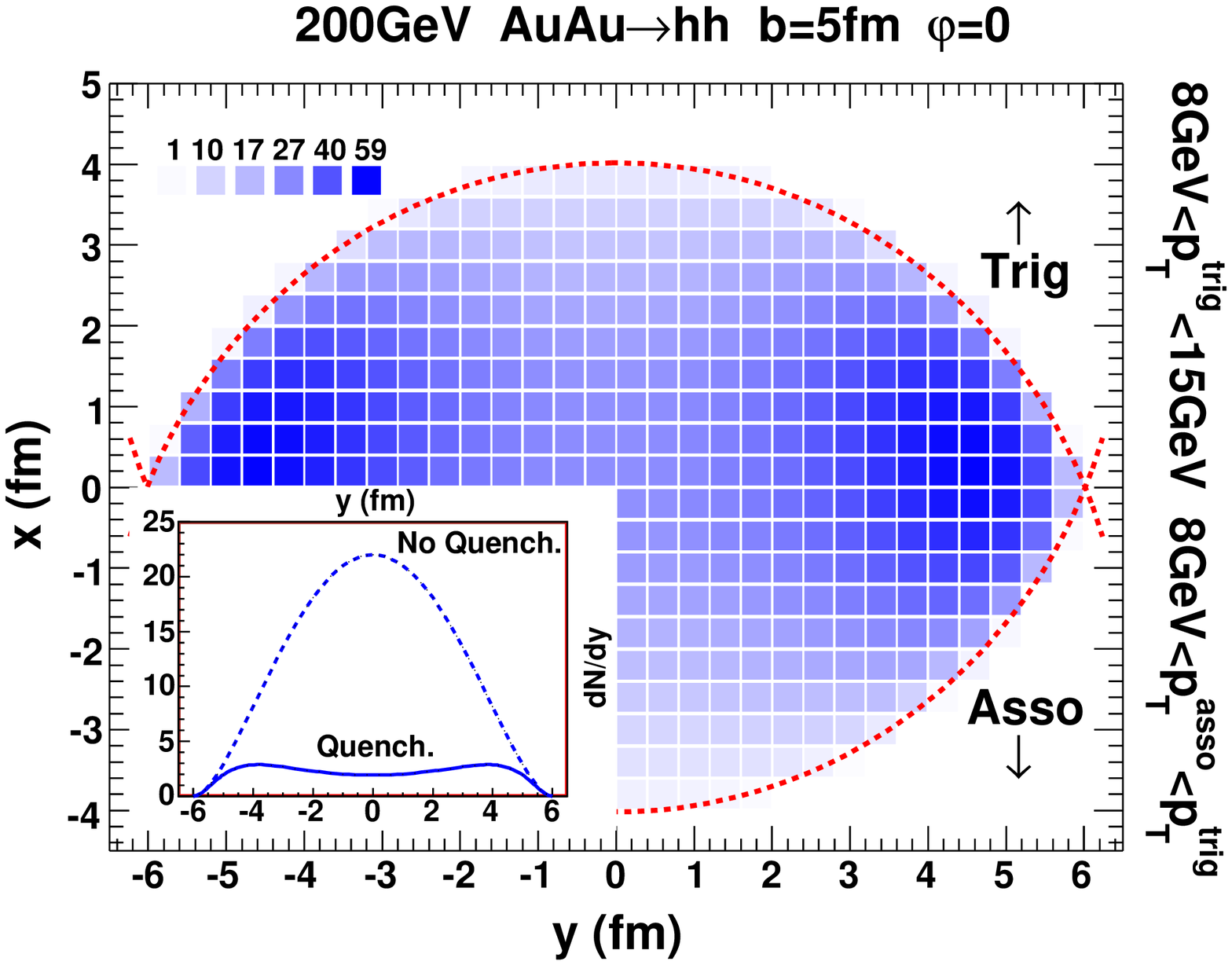}
\vskip -.15in \caption{\label{fig:contourplot} Spatial transverse
distribution (arbitrary normalization) of the produced jets
contributing the single hadron (left) along $\varphi=\pi/2$ and the
dihadron (right) along $\varphi=0$ and $\pi$. The insert is the same
distribution projected onto the $y$-axis.}
\end{figure}

By comparing NLO results with data in Fig.~\ref{fig:ppRaa} and
Fig.~\ref{fig:DaaIaa}, we get Fig.~\ref{fig:contourplot} by
choosing the factorization scale as $\mu=1.2p_T$ for single hadron
and $\mu=1.2M$ for dihadron, and the energy loss parameter
$\epsilon_0=1.68$GeV/fm in $A+A$ collisions. $M$ is the invariant
mass of the dihadron and $D_{AA}(z_T)$ was introduced in
Ref.~\cite{xn04} as a function of $z_T=p^{\rm asso}_T/p^{\rm
trig}_T$, which is the associated hadron spectrum with a triggered
hadron. The dihadron suppression factor in Fig.~\ref{fig:DaaIaa}
is given by $I_{AA}(z_T)={D_{AA}(z_T)}/{D_{pp}(z_T)}$.

For single hadron case, because of jet quenching, the dominant
contribution to the measured hadron spectra at large $p_T$ comes
from those jets that are initially produced in the outer corona of
the overlap region toward the direction of the detected (or
triggered) hadron's momentum. This is clearly illustrated in the
left plot of Fig.~\ref{fig:contourplot} by the spatial
distribution of the produced jets that have survived the
interaction with the medium and whose leading hadrons contribute
to the measured spectra. As pointed out in Ref.~\cite{eskola},
when the initial gluon density is increased such that jets
produced in the inner part of the overlapped region are completely
suppressed, the final large $p_T$ hadron production is dominated
by ``surface emission''. Therefore, the suppression factor for
single hadron spectra should never saturate but continue to
decrease with the initial gluon density as shown in the left plot
of Fig.~\ref{fig:RaaIaa}. The dependence, nevertheless, becomes
very weak when surface emission become dominant and single hadron
suppression is no longer a sensitive probe of the the initial
gluon density.

Fortrunately, in adjusting the energy loss parameter $\epsilon_0$
or the initial gluon density $\rho_0$ to fit the suppression
factors for both single hadron spectra and dihadron spectra, we
find that dihadron spectra is much more sensitive to $\epsilon_0$
than the single hadron spectra in the region $\epsilon_0=1 - 2$
GeV as shown in Fig.~\ref{fig:ppRaa}, Fig.~\ref{fig:DaaIaa} and
the left plot of Fig.~\ref{fig:RaaIaa}. One can understand this
increased sensitivity of dihadron spectra in the spatial
distribution of the dijet production that survived interaction
with the medium and contributed to the measured dihadron, as shown
in the right plot of Fig.~\ref{fig:contourplot}. Because of
trigger bias, most of the contribution comes from dijets close and
tangential to the surface of the overlapped region. However, there
are still about 25\% of the contribution coming from dijets near
the center of the overlapped region. These jets are truely
``punching'' through the medium and survived the energy loss. As
one further increases the initial gluon density, the fraction of
these punch-through jets will also vanish and the final dihadron
spectra will be dominated by the tangential jets in the outer
corona. As shown in the left plot of Fig.~\ref{fig:RaaIaa}, the
dihadron suppression factor $I_{AA}$ becomes identical to the
single hadron $R_{AA}$ and lose its sensitivity to the initial
gluon density of the medium, as would be the case with
$p_T^{trig}$=8GeV at the LHC energy. However, from a realistic
estimate of the bulk hadron production at LHC, the energy loss
parameter $\epsilon_0\approx 5$GeV/fm in central $Pb+Pb$
collisions\cite{lw02}. Of great excitements is in the right plot
of Fig.~\ref{fig:RaaIaa} that a robust $I_{AA}$ is again obtained
while the $R_{AA}$ becomes more fragile if the single and dihadron
spectra at much higher $p_T^{trig}$ (for example,
$p_T^{trig}$=20GeV) are measured in LHC.

In summary, we have used NLO pQCD parton model with effective
modified fragmentation functions due to radiative parton energy
loss to study the fragility of both single and dihadron spectra as
probes of the initial gluon density in high-energy heavy-ion
collisions. Numerical analysis shows NLO contribution behaves with
stronger quenching effect than LO contribution. Especially, we
find the high $p_T$ dihadron spectra are contributed not only by
jet pairs close and tangential to the surface of the dense matter
but also by substantial punching-through jets survived at the
center of the dense matter while the single hadron high $p_T$
spectra are only dominated by surface emission. Consequently, the
suppression factor of such high-$p_T$ hadron pairs is found to be
more sensitive to the initial gluon density than the single hadron
suppression factor.

This work was supported by DOE under contracts DE-AC02-05CH11231
and DE-FG02-97IR40122, by NSFC under project Nos. 10440420018,
10475031, 10635020, 10575043,10405011, 10447109, and by MOE of
China under projects NCET-04-0744, SRFDP-20040511005 and
CFKSTIP-704035.

\end{document}